\documentclass[onecolumn,aps,prc,showpacs,superscriptaddress]{revtex4}
\usepackage{psfig}
\newcommand {\pt}	{p_{T}}
\newcommand {\mt}	{m_{T}}

\newcommand {\psiEP}	{\psi_{\rm EP}}
\newcommand {\psiRP}	{\psi_{\rm RP}}

\newcommand {\dphi}	{\Delta\phi}
\newcommand {\cij}	{\cos2\dphi_{ij}}
\newcommand {\cphi}	{\cos2\dphi}
\newcommand {\sphi}	{\sin2\dphi}
\newcommand {\tphi}	{\tilde\phi}
\newcommand {\intpi}	{\int_{0}^{2\pi}}
\newcommand {\vtwo}[1]	{v_{2,#1}}
\newcommand {\vv}[1]	{v_2\{{\rm #1}\}}
\newcommand {\vsq}[1]	{v_2^2\{#1\}}
\newcommand {\mean}[1]	{\langle#1\rangle}
\newcommand {\acl}	{{d}}
\newcommand {\Phy}	{\mathcal{P}_{hy}}
\newcommand {\Pa}	{\mathcal{P}_{\acl}}
\newcommand {\Nhy}	{N_{hy}}
\newcommand {\Ncl}	{N_{cl}}
\newcommand {\Na}	{N_{\acl}}
\newcommand {\fcl}	{f_{\acl}}
\newcommand {\rhocl}	{\rho_{cl}}
\newcommand {\crho}	{\mean{\cos2\Delta\phi_{12}}_{\rho}}
\newcommand {\cfr}	{\mean{\cos2\Delta\phi}_{\rho}}
\newcommand {\decay}	{\rho\rightarrow\pi^+\pi^-}
\newcommand {\dNdeta}	{dN_{\rm ch}/d\eta}

\newcommand {\phid}	{\Delta\phi}
\newcommand {\etad}	{\Delta\eta}
\newcommand {\cphid}	{\cos2\Delta\phi}
\newcommand {\Nbin}	{N_{\rm bin}}
\newcommand {\be}	{\begin{equation}}
\newcommand {\ee}	{\end{equation}}
\newcommand {\bea}	{\begin{eqnarray}}
\newcommand {\eea}	{\end{eqnarray}}

\begin{document}
%%%%%%%%%%%%%%%%%%%%%%%%%%%%%%%%%%%%%%%%%%%%%%%%%%%%%%%%%%%%%%%%%%%%%%

\title{Non-flow Correlations in a Cluster Model}
\author{Quan Wang}
\author{Fuqiang Wang}
\affiliation{Department of Physics, Purdue University, 525 Northwestern Ave., West Lafayette, Indiana 47907, USA}

\begin{abstract}
We derive analytical forms for non-flow contributions from cluster correlation to two-particle elliptic flow ($\vv{2}$) measure. We estimate non-flow contribution from $\decay$ decays and find it is negative but not a major contributor to non-flow effect in $\vv{2}$. We also estimate non-flow contribution from the recent STAR measurement of two-particle angular correlations.
\end{abstract}


\maketitle

%%%%%%%%%%%%%%%%%%%%%%%%%%%%%%%%%%%%%%%%%%%%%%%%%%%%%%%%%%%%%%%%%%%%%%
\section{Introduction}

Azimuthal distribution of charged hadrons in non-central heavy-ion collisions is anisotropic~\cite{flow}. Expressed in Fourier series, the leading anisotropic term is the second harmonic, called elliptic flow ($v_2$). Anisotropies may stem from early stage hydrodynamic expansion (hydro-flow) of the bulk medium created in relativistic heavy-ion collisions, thus making their measurement interesting~\cite{whitepaper}. However, other mechanisms, of non-flow nature, such as jet-correlation and resonance decays, can play a significant role~\cite{nonflow}. A significant effort of anisotropic flow studies at RHIC is to investigate the magnitude of non-flow effects.

There are several methods to measure $v_2$~\cite{v2method}; all of them are affected by non-flow with differing degrees of sensitivity. One method, called the event plane method, is to construct the event plane $\psiEP$ from all charged particles except those of interest (exploiting the very fact that particles are anisotropically distributed) and calculate $\vv{RP}=\mean{\cos2(\phi-\psiEP)}/\mean{\cos2\Delta\psiEP}$ for particles of interest, where $\mean{\cos2\Delta\psiEP}$ is the event plane resolution. This method is affected by non-flow in the interested particles as well as in all other particles used in the event plane construction. Another method, called two-particle method, is to calculate $\vsq{2}=\mean{\cphi}$ using all particle pairs of interest where $\dphi$ is pair opening azimuthal angle. This method is affected by non-flow in the interested particles. It can be shown that the $\vv{EP}$ and $\vv{2}$ are approximately equal~\cite{v2method,trainor}. The third method, called four-particle method, is to obtain $\vv{4}$ from four-particle cumulant~\cite{v4}. This method is less affected by non-flow from particle clustering because the number of clusters with more than four particles is reduced and because non-flow effect is diluted by particle multiplicity to a high power. This method is not affected by resonance decays. On the other hand, flow fluctuations, mainly due to initial geometry eccentricity fluctuations, have different effects on $\vv{2}$ ($\vv{EP}$) and $\vv{4}$~\cite{Voloshin_fluc,PHOBOS}. The fluctuation effect is positive in $\vv{2}$ but negative in $\vv{4}$. The difference between $\vv{2}$ and $\vv{4}$ is therefore a net effect of non-flow and flow fluctuations.

In this paper we investigate non-flow effect originating from clusters. We focus on only $\vv{2}$ for two primary reasons: one is that $\vv{2}$ is related to two-particle azimuthal correlation simply by $\vsq{2}=\mean{\cphi}$, and the other is that non-flow effect in $\vv{2}$ comes from only particles used in the correlation so it is relatively straightforward to disentangle.

\section{Non-flow effect from cluster correlations}

Suppose an event is composed of particles from hydro-medium and clusters of various types (such as minijets and resonance decays). Particle pairs can be decomposed into four sources: 
\begin{itemize}
\item particle pairs from hydro-medium ($B$),
\item particle pairs from same cluster ($C$),
\item particle pairs between hydro-medium and clusters ($X$), and
\item particle pairs between clusters ($Y$).
\end{itemize}
The total sum of the cosines of pair opening angles is
\be
\sum_{i\neq j}\cij=B+\sum_{k\in cluster}C+\sum_{k\in cluster}2X+\sum_{(k_1\neq k_2)\in cluster}Y,
\label{eq1}
\ee
where
\bea
B&=&\sum_{(i\neq j)\in hydro}\cij,\\
C&=&\sum_{(i\neq j)\in k}\cij,\label{eq:C}\\
X&=&\sum_{i\in k}\sum_{j\in hydro}\cij,\label{eq:X}\\
Y&=&\sum_{i\in k_1}\sum_{j\in k_2}\cij.\label{eq:Y}
\eea
Here $i,j$ are particle indices, $\dphi_{ij}=\phi_i-\phi_j$, and $k$ stands for a cluster. Below we derive analytical form for each source.

\subsection{Background flow correlation}

Hydro-background particle correlation is only from hydrodynamic anisotropic flow:
\be
B=\sum_{(i\neq j)\in hydro}\cij=\Phy\mean{\cphi_{hy}}=\Phy\vsq{2}_{hy}
\ee
where $\Phy=\mean{\Nhy(\Nhy-1)}$ is the number of background pairs.

\subsection{Particle correlation within cluster}

Particle correlation within cluster is given by
\be
C=\sum_{(i\neq j)\in k}\cij=\intpi\Pa(\tphi_k)\rhocl(\tphi_k)d\tphi_k\intpi \fcl(\dphi_i,\tphi_k)d\dphi_i\intpi \fcl(\dphi_j,\tphi_k)d\dphi_j\cos2(\dphi_i-\dphi_j).
\ee
Here $\dphi_{i,j}=\phi_{i,j}-\phi_k$ is the azimuthal angles of particles in cluster $k$ relative to cluster axis $\phi_k$ (which can be defined just for convenience); $\fcl(\dphi,\tphi_k)$ is the correlation function of (daughter) particles inside cluster $k$ relative to the cluster axis $\phi_k$, generally dependent of the cluster axis $\tphi_k=\phi_k-\psiRP$ relative to the reaction plane, and $\intpi \fcl(\dphi,\tphi_k)\dphi\equiv1$; $\Pa(\tphi_k)$ is number of (daughter) particle pairs in cluster $k$, generally dependent of the cluster axis; $\rhocl(\tphi_k)$ is the density function of cluster $k$ relative to the reaction plane, which we will assume is given by elliptic flow of clusters, and $\intpi\rhocl(\tphi_k)d\tphi_k\equiv1$. Note, there can be many types of clusters (e.g. jet-correlation, resonance decays); the subscript `$cl$' stands for one type of clusters and we have omitted summation over all types of clusters from the formulism; in this work we will discuss only one type of clusters at a time. 

In general,
\be
C=\Pa\mean{\cij}_{cl}
\ee
where $\Pa$ is average number of pairs per cluster and $\mean{\cij}_{cl}$ is the average cosine of twice pair opening angle in the cluster. 

If particles inside cluster are independent of each other except all of them are correlated with the cluster axis, then we can factorize the correlation terms and obtain
\bea
C&=&\intpi\Pa(\tphi_k)\rhocl(\tphi_k)d\tphi_k\left[\left(\intpi \fcl(\dphi,\tphi_k)\cphi d\dphi\right)^2+\left(\intpi \fcl(\dphi,\tphi_k)\sphi d\dphi\right)^2\right]\nonumber\\
&=&\intpi\Pa(\tphi_k)\left(\mean{\cphi}_{\tphi_k}^2+\mean{\sphi}_{\tphi_k}^2\right)\rhocl(\tphi_k)d\tphi_k.
\eea
Here $\mean{\cphi}_{\tphi_k}$ and $\mean{\sphi}_{\tphi_k}$ $(\dphi=\phi-\phi_k)$ are averages within cluster $k$, and are generally dependent of the cluster axis $\tphi_k$. 

If the cluster correlation function $\fcl(\dphi,\tphi_k)$ is symmetric about $\dphi=0$, then $\mean{\sphi}_{\tphi_K}=0$. Further, in the special case where particle correlation in clusters does not vary with cluster location $\tphi_k$, i.e., $\Pa={\rm const.}$ and $\fcl(\dphi,\tphi_k)={\rm const.}$, then
\be
C=\Pa\mean{\cphi}_{cl}^2.
\ee

\subsection{Background-cluster correlation}

Correlation between cluster particles and hydro-medium particles is given by
\bea
X=\sum_{i\in k}\sum_{j\in hydro}\cij&=&\intpi \Nhy(\tphi_{k=1,2,...})\rho_{hy}(\tphi_{hy})d\tphi_{hy}\times\nonumber\\
&&\intpi\rhocl(\tphi_k)d\tphi_k\intpi\Na(\tphi_k)\fcl(\dphi_i,\tphi_k)d\dphi_i\left[\cos2(\phi_i-\phi_{hy})\right].
\eea
Here $\tphi_{hy}=\phi_{hy}-\psiRP$, and $\rho_{hy}(\tphi_{hy})$ is the density function of hydro particles relative to the reaction plane (i.e., anisotropic hydro flow); $\Na(\tphi_k)$ is number of (daughter) particles in cluster and is generally dependent of the cluster axis $\tphi_k$. For generality we have taken the number of hydro-medium particles $\Nhy(\tphi_{k=1,2...})$  to depend on positions of all clusters. Such dependence can arise in real data analysis, such as jet-correlation analysis, from interplay between centrality cut and biases due to selection of specific clusters. Rewriting $\phi_i-\phi_{hy}=\dphi_i+\tphi_k-\tphi_{hy}$, we have
\bea
X&=&\intpi \Nhy(\tphi_{k=1,2,...})\rho_{hy}(\tphi_{hy})\cos2\tphi_{hy}d\tphi_{hy}\times\nonumber\\
&&\intpi\rhocl(\tphi_k)d\tphi_k\intpi \Na(\tphi_k)\fcl(\dphi_i,\tphi_k)\cos2(\dphi_i+\tphi_k)d\dphi_i.
\eea
Here we have used $\displaystyle{\intpi}d\tphi_k\rhocl(\tphi_k)\displaystyle{\intpi}d\dphi_k\fcl(\dphi_k,\tphi_k)\sin2(\dphi_k+\tphi_k)=0$ because of symmetries $\fcl(\dphi_k,\tphi_k)=\fcl(-\dphi_k,-\tphi_k)$ and $\rhocl(\tphi_k)=\rhocl(-\tphi_k)$. Note, due to elliptic flow of clusters, cluster particles acquire elliptic flow
\be
\vtwo{\acl}\equiv\mean{\cos2(\phi-\psiRP)}=\frac{1}{\Na}\intpi\rhocl(\tphi_k)d\tphi_k\intpi \Na(\tphi_k)\fcl(\dphi_i,\tphi_k)\cos2(\dphi_i+\tphi_k)d\dphi_i.
\label{eq6}
\ee
Using the notation in Eq.~(\ref{eq6}), we have
\be
X=\Nhy\Na \vtwo{hy}\vtwo{\acl}=\Nhy\Na\vv{2}_{hy}\vv{2}_{\acl}.
\ee
Here the product of the $v_2$'s includes flow fluctuation, and equals to the product of two-particle $v_2$'s. This is because $\vv{2}$ of hydro-particles and cluster particles contain only fluctuation; non-flow does not exist between hydro-particles, nor between particles from different clusters. (Note, two `clusters' can originate from a common ancestor, such as jet fragmentation into two $\rho$ mesons which in turn decay into two pairs of pions. In our formulism, such `clusters' are considered to be parts of a single cluster rather than two $\rho$-decay clusters.)

Again, in the special case where particle correlation in clusters does not vary with cluster location $\tphi_k$ ($\Na={\rm const.}$, $\Nhy={\rm const.}$, and $\fcl(\dphi,\tphi_k)=\fcl(\dphi)$), Eq.~(\ref{eq6}) becomes
\be
\vtwo{\acl}=\intpi\rhocl(\tphi_k)d\tphi_k\cos2\tphi_k\intpi\fcl(\dphi_i)\cphi_id\dphi_i=\vtwo{cl}\mean{\cphi}_{cl},
\label{eq8}
\ee
and we have
\be
X=\Nhy\Na\vv{2}_{hy}\vv{2}_{cl}\mean{\cphi}_{cl},
\ee

\subsection{Particle correlation between clusters}

Correlation between particles from different clusters is given by
\bea
Y=\sum_{i\in k_1}\sum_{j\in k_2}\cij&=&\intpi\rhocl(\tphi_{k_1})d\tphi_{k_1}\intpi \Na(\tphi_{k_1})\fcl(\dphi_i,\tphi_{k_1})d\dphi_i\times\nonumber\\
&&\intpi\rhocl(\tphi_{k_2})d\tphi_{k_2}\intpi \Na(\tphi_{k_2})\fcl(\dphi_j,\tphi_{k_2})d\dphi_j\left[\cos2(\phi_i-\phi_j)\right],
\eea
where $k_1$ and $k_2$ stand for two clusters. Rewriting $\phi_i-\phi_j=\dphi_i+\tphi_{k_1}-\dphi_j-\tphi_{k_2}$, we obtain
\be
Y=\Na^2v^2_{2,\acl}=\Na^2\vsq{2}_{\acl},
\label{eq10}
\ee
where $\vtwo{\acl}$ is given by Eq.~(\ref{eq6}). Again the cluster particle elliptic flow squared in Eq.~(\ref{eq10}) contains flow fluctuation.

In the special case where particle correlation in clusters does not vary with cluster location $\tphi_k$, we have
\be
Y=\Na^2\vsq{2}_{cl}\mean{\cphi}^2_{cl}.
\ee

\subsection{Summary of non-flow effect from cluster correlations}

To summarize, let us now obtain the relationship between two-particle elliptic flow $\vv{2}$ that is affected by non-flow, and the real hydro-type two-particle elliptic flow $\vv{2}_{hy}$. Assuming Poisson statistics, Eq.~(\ref{eq1}) gives
\bea
N^2\vsq{2}&=&N^2_{hy}\vsq{2}_{hy}+\Ncl\Na^2\mean{\cij}_{cl}+2\Nhy\Ncl\Na\vv{2}_{hy}\vv{2}_{\acl}+\Ncl(\Ncl-1)\Na^2\vsq{2}_{\acl}\nonumber\\
&=&\left(\Nhy\vv{2}_{hy}+\Ncl\Na\vv{2}_{\acl}\right)^2+\Ncl\Na^2\left(\mean{\cij}_{cl}-\vsq{2}_{\acl}\right),
\label{eq11}
\eea
where $N=\Nhy+\Ncl\Na$, $\Ncl$ is average number of clusters, and we have taken distributions of total multiplicity and number of particles per cluster to be Poisson, so that $\mean{N(N-1)}=N^2$ and $\mean{\Na(\Na-1)}=\Na^2$. We have taken the number of cluster pairs to be $\Ncl(\Ncl-1)$ (i.e., not Poisson) so that the total number of pairs adds up to $N^2$~\cite{note_Poisson}. 
Rearranging, we have
\be
\vsq{2}=\left(\frac{\Nhy}{N}\vv{2}_{hy}+\frac{\Ncl\Na}{N}\vv{2}_{\acl}\right)^2+\frac{\Ncl\Na^2}{N^2}\left(\mean{\cij}_{cl}-\vsq{2}_{\acl}\right).
\label{eq12}
\ee
For many cluster types, Eq.~(\ref{eq12}) is generalized to
\be
\vsq{2}=\left(\frac{\Nhy}{N}\vv{2}_{hy}+\sum_{cl}\frac{\Ncl\Na}{N}\vv{2}_{\acl}\right)^2+\sum_{cl}\frac{\Ncl\Na^2}{N^2}\left(\mean{\cij}_{cl}-\vsq{2}_{\acl}\right).
\ee

We shall focus on the special case where all clusters are of the same type and particle correlation in clusters does not vary with cluster axis relative to the reaction plane. Using Eq.~(\ref{eq8}), Eq.~(\ref{eq12}) becomes
\be
\vsq{2}=\left(\frac{\Nhy}{N}\vv{2}_{hy}+\frac{\Ncl\Na}{N}\vv{2}_{cl}\mean{\cphi}_{cl}\right)^2+\frac{\Ncl\Na^2}{N^2}\left(\mean{\cij}_{cl}-\vsq{2}_{cl}\mean{\cphi}_{cl}^2\right).
\label{eq14}
\ee
Eq.~(\ref{eq14}) can be rewritten into
\bea
\vsq{2}&=&\vsq{2}_{hy}+2\frac{\Nhy}{N}\frac{\Ncl\Na}{N}\vv{2}_{hy}\left(\vv{2}_{cl}\mean{\cphi}_{cl}-\vv{2}_{hy}\right)+\nonumber\\
&&\left(\frac{\Ncl\Na}{N}\right)^2\left(\vsq{2}_{cl}\mean{\cphi}_{cl}^2-\vsq{2}_{hy}\right)+\frac{\Ncl\Na^2}{N^2}\left(\mean{\cij}_{cl}-\vsq{2}_{cl}\mean{\cphi}_{cl}^2\right).
\label{eq15}
\eea
The second term on the r.h.s.~is non-flow (beyond hydro-flow) due to correlation between hydro-particles and cluster particles in excess of that between two hydro-particles, and the third term is that due to correlation between particles from different clusters. These non-flow contributions, which are beyond hydro-flow, can be positive or negative, depending on the relative magnitudes of background particle flow and cluster flow diluted by particle spread inside cluster. The non-flow contributions are positive when $\vv{2}_{cl}\mean{\cphi}_{cl}>\vv{2}_{hy}$ and negative when $\vv{2}_{cl}\mean{\cphi}_{cl}<\vv{2}_{hy}$. This can be easily understood because if $\vv{2}_{cl}=\vv{2}_{hy}$ , then the angular smearing of particles inside each cluster, $\mean{\cphi}_{cl}$, makes the angular variation of cluster particles less than that of hydro-particles, resulting in a negative non-flow contribution. If the net effect of cluster anisotropy and particle distribution inside clusters, $\vv{2}_{cl}\mean{\cphi}_{cl}$, equals to hydro anisotropy, then cluster particles and hydro-particles have the same angular variation relative to the reaction plane, resulting in zero non-flow from cross-pairs between hydro-particles and cluster particles and between particles from different clusters.

%Likely $\vsq{2}_{cl}<<1$, so it can be neglected from the last term. 
The second part of the last term of Eq.~(\ref{eq15}) r.h.s., $\frac{\Ncl\Na^2}{N^2}\vsq{2}_{cl}\mean{\cphi}_{cl}^2$, arises from assumptions of Poisson statistics, and can be safely neglected because generally $\vsq{2}_{cl}<<1$. The first part of the last term of Eq.~(\ref{eq15}) r.h.s., $\frac{\Ncl\Na^2}{N^2}\mean{\cij}_{cl}$, is non-flow due to correlation between particles in the same cluster. This non-flow contribution can also be positive or negative. If particle emissions within clusters are independent, $\mean{\cij}_{cl}=\mean{\cphi}_{cl}^2$ , then Eq.~(\ref{eq15}) becomes
\bea
\vsq{2}&=&\vsq{2}_{hy}+2\frac{\Nhy}{N}\frac{\Ncl\Na}{N}\vv{2}_{hy}\left(\vv{2}_{cl}\mean{\cphi}_{cl}-\vv{2}_{hy}\right)+\nonumber\\
&&\left(\frac{\Ncl\Na}{N}\right)^2\left(\vsq{2}_{cl}\mean{\cphi}_{cl}^2-\vsq{2}_{hy}\right)+\frac{\Ncl\Na^2}{N^2}\left(1-\vsq{2}_{cl}\right)\mean{\cphi}_{cl}^2.
\label{eq16}
\eea
In this case the non-flow contribution due to particle correlation within clusters %(the last term in Eq.~(\ref{eq16}) r.h.s.) 
can only be positive.

We note that the non-flow contributions from the second and third term of Eq.~(\ref{eq15}) r.h.s.~have the identical azimuthal shape relative to the reaction plane as that of hydro-flow, because they arise from the common correlation of clusters and hydro-particles to the reaction plane. As a result these non-flow contributions will unlikely be separated from medium hydro-flow in inclusive measurement of azimuthal correlation. To separate these two contributions, one needs to identify clusters and measure two-cluster azimuthal correlation. In fact, elliptic flow is often defined as the second harmonic of particle distribution relative to the reaction plane, $\vv{RP}=\mean{\cos2(\phi-\psiRP)}$ . For events composed of hydro-particles and clusters, we have
\be
\vv{RP}=\frac{N_{hy}}{N}\mean{\cos2(\phi-\psiRP)}_{hy}+\frac{\Ncl\Na}{N}\mean{\cos2(\phi-\psiRP)}_{cl}=\frac{N_{hy}}{N}\vv{RP}_{hy}+\frac{\Ncl\Na}{N}\vv{RP}_{cl}\mean{\cphi}_{cl}.
\label{eq17}
\ee
This is analogous to the terms in the first pair of parentheses on Eq.~(\ref{eq14}) r.h.s.~except the latter contains flow fluctuation. The elliptic flow definition by Eq.~(\ref{eq17}) contains cluster contribution through angular spread of particles in clusters, $\mean{\cphi}_{cl}$, and anisotropy of the clusters themselves, $\vv{RP}_{cl}$. This raises question to comparisons often made between elliptic flow measurements and hydro calculations which may include flow fluctuation but does not include cluster correlations. 

\section{Estimate of non-flow from two-body resonance decays}

In relativistic heavy-ion collisions, a large fraction of final state pions come from $\rho$ decays. Charged pion pairs from $\decay$ decays have intrinsic angular correlation due to decay kinematics. In this section, we estimate the non-flow effect from this intrinsic angular correlation on pion elliptic flow. 

Suppose a parent particle of mass $M$ decays into two daughter particles of equal mass $m$ with decay angle $\theta$ (the angle a daughter particle makes in the parent c.m. frame with the parent direction of motion). The opening angle between the two daughters is straightforward to calculate: 
\be
\cos2\Delta\theta_{12}=1-\frac{8\gamma^2\beta^2(1-4\alpha^2)\sin^2\theta}{\left[\gamma^2+4\alpha^2-\gamma^2\beta^2(1-4\alpha^2)\cos^2\theta\right]^2-16\gamma^2\alpha^2},
\ee
where $\beta$ is the parent speed, $\gamma=1/\sqrt{1-\beta^2}$ its Lorentz factor, and $\alpha=m/M$. To obtain the opening angle projected onto the transverse plane, however, is a bit tedious. So we resort to MC sampling of $\decay$ decays. We assume uniform rapidity distribution for $\rho$ within $|y|<2$. We calculate the average cosine of twice the opening azimuthal angle, $\crho$, between the two daughter pions in the lab frame if both are within rapidity $|y|<1$. We assume isotropic decay in $\rho$ c.m.~frame~\cite{note}. 

Figure 1 (left panel) shows $\crho$ as a function of $\rho$ transverse momentum ($\pt$). $\crho=1$ for $\pt=0$, because the decay pions are back-to-back in the lab frame. $\crho$ drops quickly with $\pt$ due to closing of pion pairs from boosting. $\crho$ becomes negative at intermediate $\pt$ when the opening angle is narrowed towards $\pi/2$. When $\pt$ increases further, strong boost focusing narrows the pair opening angle to asymptotically approaching zero.

\begin{figure*}[hbt]
\centerline{
\psfig{file=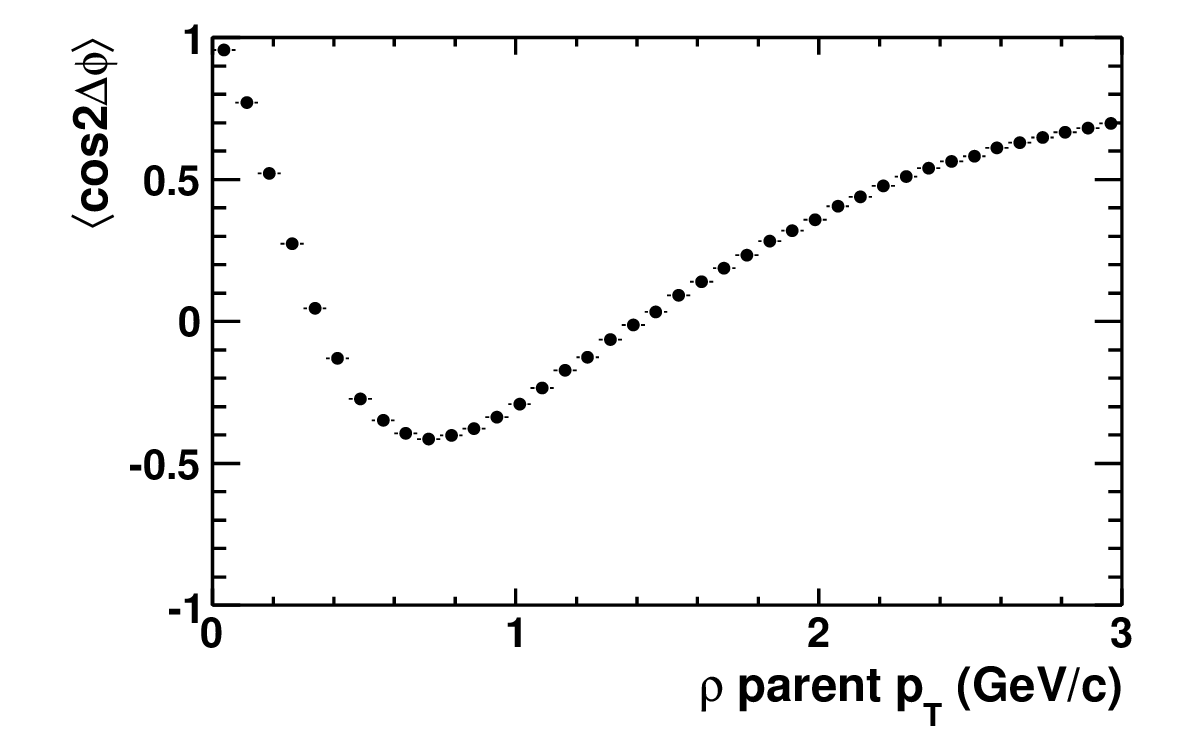,width=0.5\textwidth}
\psfig{file=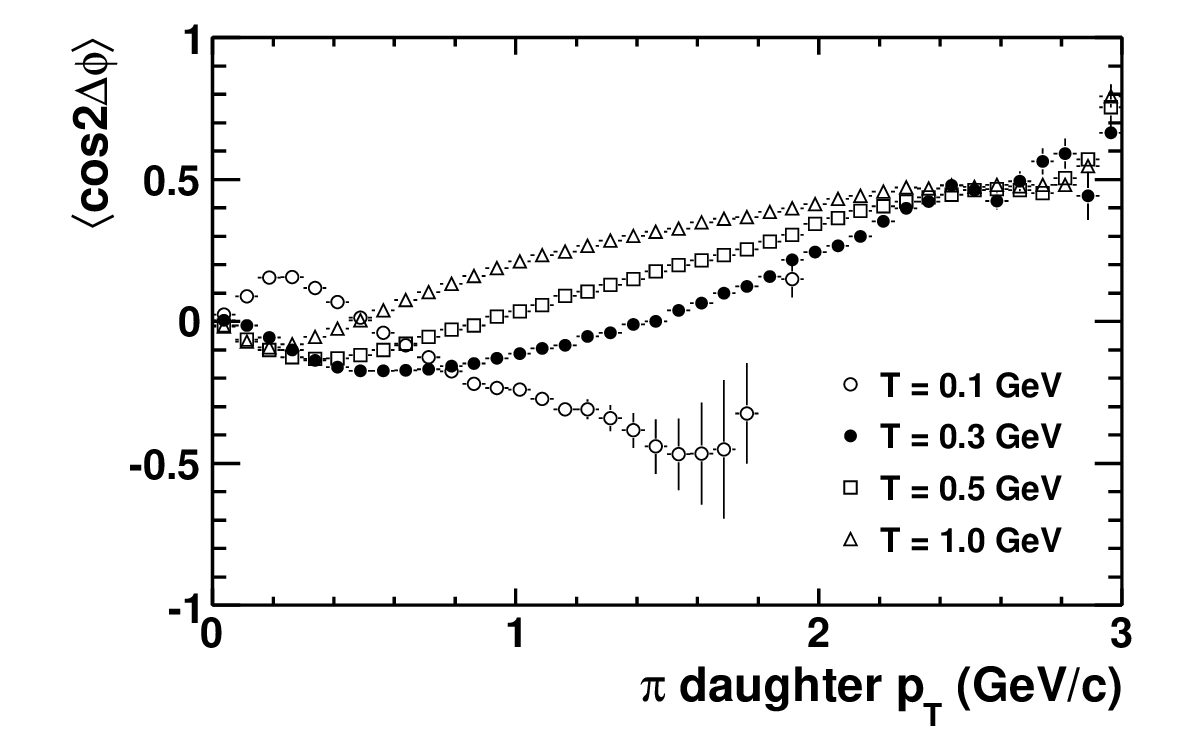,width=0.5\textwidth}
}
\caption{Left panel: $\crho$ vs parent $\rho$-meson $\pt$. Flat $\pt$ distribution is used to have uniform statistics. Right panel: $\crho$ vs daughter pion $\pt$. Each decay is entered twice in this plot, at $\pt$ of each pion. The transverse distribution is assumed to be $\mt$-exponential with inverse slope $T$. Rapidity range $|y|<2$ is simulated for $\rho$ and daughter pions are restricted within $|y|<1$ for both panels. Uniform decay angle distribution is used. One million decays are simulated for each set of data points.}
\label{fig}
\end{figure*}

To calculate the average $\crho$ over all $\pt$, we take exponential in $\mt=\sqrt{M^2+\pt^2}$ for transverse distribution as measured in Au+Au collisions~\cite{rho}:
\be
\frac{d^2N}{\mt d\mt dy}\propto\exp\left(-\frac{\mt}{T}\right).
\ee
Figure 1 (right panel) shows $\crho$ as a function of daughter pion $\pt$. Each decay is entered twice in this plot, at $\pt$ of each pion. Table I lists the overall $\crho$ versus $T$. $\crho$ decreases and then increases with $T$. In the range $T=0.2-0.5$~GeV relevant for heavy-ion collision~\cite{rho}, $\crho$ is negative and relatively constant. Emissions of the two daughter particles are not independent, so $\crho\neq\cfr^2$, where $\dphi_{12}$ is azimuthal opening angle of the daughter pion pair and $\dphi$ is azimuthal angle of daughter pion relative parent direction of motion. Table I lists $\cfr$ together with $\crho$ at various $T$ values.

\begin{table*}
\caption{$\decay$ decay $\crho$ versus $T$ ($\rho$ $\mt$ exponential inverse slope).}
\label{tab1}
\begin{tabular}{cccccccc}
\hline\hline
$T$ (GeV) & 0.1 & 0.2 & 0.3 & 0.4 & 0.5 & 0.7 & 1.0 \\
\hline
$\crho$ & $0.13$ & $-0.13$ & $-0.17$ & $-0.14$ & $-0.10$ & $-0.03$ & $0.05$ \\
$\mean{\cos2\dphi}_{\rho}$ & $0.07$ & $0.18$ & $0.28$ & $0.36$ & $0.41$ & $0.49$ & 0.55\\
\hline\hline
\end{tabular}
\end{table*}

Let $x$ be the fraction of final state pions from $\rho$ decays; the number of $\rho$'s (clusters) is $xN/2$. We will assume Poisson statistics except for the number of daughters ($N_k=2$) and daughter pairs ($N_k(N_k-1)=2$) per decay that are both fixed. Because of the non-Poisson statistics of the $\rho$-decay daughters, Eq.~(\ref{eq15}) cannot be readily applied. However, this can be easily revamped by noting that the individual non-flow contributions in Eq.~(\ref{eq1}) are now:
\bea
B&=&(1-x)^2N^2\vsq{2}_{hy},\\
C&=&2\crho,\\
X&=&2(1-x)N\vv{2}_{hy}\vv{2}_{\rho}\cfr,\\
Y&=&4\vsq{2}_{\rho}\cfr^2.
\eea
Thus Eq.~(\ref{eq1}) becomes
\be
N^2\vsq{2}=(1-x)^2N^2\vsq{2}_{hy}+xN\crho+2x(1-x)N^2\vv{2}_{hy}\vv{2}_{\rho}\cfr+xN(xN-1)\vsq{2}_{\rho}\cfr^2,
\ee
where we have taken the number of cluster pairs to be $\left(\frac{xN}{2}\right)\cdot\left(\frac{xN-1}{2}\right)$ so that the total number of particle pairs adds up to $N^2$. Finally we obtain
\bea
\vsq{2}&=&\vsq{2}_{hy}+2x\vv{2}_{hy}\left(\vv{2}_{\rho}\cfr-\vv{2}_{hy}\right)+\nonumber\\
&&x^2\left(\vv{2}_{\rho}\cfr-\vv{2}_{hy}\right)^2+\frac{x}{N}\left(\crho-\vsq{2}_{\rho}\cfr^2\right)\nonumber\\
&=&\left[(1-x)\vv{2}_{hy}+x\vv{2}_{\rho}\cfr\right]^2+\frac{x}{N}\left(\crho-\vsq{2}_{\rho}\cfr^2\right).
\label{eq18}
\eea
%\be
%\vsq{2}=\left[(1-x)\vv{2}_{hy}+x\vv{2}_{\rho}\cfr\right]^2+\frac{x}{N}\left(\crho-\vsq{2}_{\rho}\cfr^2\right).
%\label{eq18}
%\ee
The second and third terms of Eq.~(\ref{eq18}) r.h.s.~are non-flow contributions from correlations between decay particles and hydro-particles and between daughter particles from different decays. Obviously, the magnitude of these non-flow depends on $\vv{2}_{\rho}$: if $\vv{2}_{\rho}\cfr>\vv{2}_{hy}$, then their non-flow is positive, and otherwise, negative. The last term of Eq.~(\ref{eq18}) r.h.s.~is the non-flow contribution from correlation between the two decay daughters, and is negative for $\decay$ decays with our chosen kinematics. Since $\rho$ is short-lived, $\rho$ and hydro-medium pions are in detailed balance at the early stage of the collision where elliptic flow is generated. It is therefore reasonable to assume that $\vv{2}_{\rho}\cfr=\vv{2}_{hy}$, i.e., $\rho$ decay pions and hydro-medium pions possess the same elliptic flow and are indistinguishable. In this case the non-flow contribution is entirely from correlations between daughter particles of the same decay:
\be
\vsq{2}=\vsq{2}_{hy}+\frac{x}{N}\left(\crho-\vsq{2}_{\rho}\cfr^2\right)\approx v^2_{2,hy}+\frac{x}{N}\crho.
\ee
Here we have made assumption $\vsq{2}_{\rho}<<1$. It is worth to note that only the final-stage $\rho$ decays are relevant because pions from decays at early times rescatter and lose their intrinsic angular correlation. Those decay pions will eventually become part of the hydrodynamic medium after rescattering. The fraction of $\rho$-decay pions $x$, therefore, should refer to only those final-stage $\rho$ decays that are experimentally measured.

Table II lists our estimates of non-flow contributions from $\decay$ decays for various centrality bins together with $v_2$ measurements~\cite{v2data}. The $\rho$ $\mt$ exponential inverse slope is taken to be $T=0.3$~GeV as measured in peripheral Au+Au collisions~\cite{rho} and consequently $\crho=-0.17$; other $T$ values reasonable for heavy-ion collision do not alter $\crho$ significantly. The fraction of decay pions is taken to be $x=40\%$ as measured in peripheral Au+Au collisions~\cite{rho}; its centrality dependence is weak as the chemical freeze-out temperature is measured to be independent of centrality~\cite{spectra} and losses due to final state rescattering are likely offset by regeneration processes~\cite{kstar}. The listed $\vv{2}_{hy}$ values are calculated from the measured $\vv{2}$ assuming the only non-flow contribution is from $\decay$ decays and $\vv{2}_{\rho}\cfr=\vv{2}_{hy}$. Magnitude of non-flow contribution is calculated by $1-\frac{\vv{2}_{hy}}{\vv{2}}$, and is typically a few percent for most centrality bins. This is small (and with wrong sign) compared to the difference between the $\vv{2}$ and $\vv{4}$ measurements (see $1-\frac{\vv{4}}{\vv{2}}$ in Table II). Note that this non-flow contribution is only from correlation between daughter particles from the same decay, whilst contributions from correlations between particles from different decays and between decay particles and medium hydro-particles are arguably small (i.e., $\vv{2}_{\rho}\cfr=\vv{2}_{hy}$) and are neglected here. Even not small, those contributions should be already included in the $\vv{4}$ measurement. Therefore, the large difference between the measured $\vv{2}$ and $\vv{4}$ suggests that resonance decays are not a major contributor to non-flow at RHIC. The difference must then come from other major non-flow effects and/or flow fluctuations resulting from initial eccentricity fluctuations.

\begin{table*}
\caption{Estimates of non-flow contributions from $\decay$ decays in Au+Au collisions at 200 GeV as a function of centrality together with two-particle $\vv{2}$ and four-particle $\vv{4}$ measurements~\cite{v4}. The $\dNdeta$ results are from Ref.~\cite{Levente}. The $\rho$ transverse distribution is taken to be $\mt$ exponential with an inverse slope of 0.3 GeV~\cite{rho} (and consequently $\crho=-0.17$), and the fraction of pions from $\decay$ decays is taken to be 40\%~\cite{rho}, both independent of centrality. It is also assumed that decay pions and bulk hydro-medium pions possess the same elliptic flow, i.e., $\vv{2}_{\rho}\cfr=\vv{2}_{hy}$, and are indistinguishable. $\vv{2}_{hy}$ is calculated from the measured $\vv{2}$ assuming the only non-flow contribution is from $\decay$ decays. Relative non-flow contribution from resonance decays is calculated as $1-\frac{\vv{2}_{hy}}{\vv{2}}$, to be compared to non-flow effect indicated by $1-\frac{\vv{4}}{\vv{2}}$ from measurement.}
\label{tab2}
\begin{tabular}{cccccccc}
Centrality & $\dNdeta$ & $\vv{2}$ & $\vv{4}$ & $-\sqrt{-\frac{x}{N}\crho}$ & $\vv{2}_{hy}$ & $1-\frac{\vv{2}_{hy}}{\vv{2}}$ & $1-\frac{\vv{4}}{\vv{2}}$ \\
\hline\hline
  0-5\% & 691 & 2.41\% &	& $-0.70$\% & 2.51\% & $-4.15$\%	 \\
 5-10\% & 558 & 3.55\% & 2.53\% & $-0.78$\% & 3.63\% & $-2.39$\% & 28.73\% \\
10-20\%	& 421 & 4.97\% & 4.27\% & $-0.90$\% & 5.05\% & $-1.62$\% & 14.08\% \\
20-30\% & 287 & 6.42\% & 5.66\% & $-1.09$\% & 6.51\% & $-1.43$\% & 11.84\% \\
30-40\% & 195 & 7.29\% & 6.33\% & $-1.32$\% & 7.41\% & $-1.63$\% & 13.17\% \\
40-50\% & 126 & 7.64\% & 6.43\% & $-1.64$\% & 7.81\% & $-2.29$\% & 15.84\% \\
50-60\% &  78 & 7.59\% & 6.18\% & $-2.09$\% & 7.87\% & $-3.71$\% & 18.58\% \\
60-70\% &  45 & 7.25\% & 5.68\% & $-2.75$\% & 7.75\% & $-6.95$\% & 21.66\% \\
70-80\% &  22 & 6.88\% &	& $-3.93$\% & 7.92\% & $-15.17$\% \\
\hline\hline
\end{tabular}
\end{table*}

\section{Non-flow from ``minijet'' correlations}

Strong angular correlations in $(\etad,\phid)$ between soft particles have been observed by STAR~\cite{minijet}. They reveal in part characteristics of jet-like correlations from small energy partons (the so-called minijets), such as the narrow $\phid$ correlations. On the other hand, the correlations are measured to be extraordinarily wide in $\etad$. The long $\etad$ correlation ridge may be due entirely to the medium, such as fluctuations of color flux tubes in the initial stage of the collisions. The ridge particles are then focused into the narrow $\phid$ window by the radial flow boost, a result of the medium's hydrodynamic expansion. This would suggest that the measured two-particle angular correlations may not necessarily be due to jet-like correlations, but hydrodynamic expansion of the medium. Even if the measured angular correlations are indeed due to jet-like correlations, those correlations must have been influenced significantly by the medium hydrodynamic flow. It is, therefore, difficult to separate flow and non-flow.

The measured angular correlation raw data (i.e., pair density per charged hadron) were decomposed into two components: elliptic flow and minijet correlations. The latter is assumed to consist of a two-dimensional Gaussian in $(\etad,\phid)$ and a dipole moment in $\dphi$:
\be
f(\phid,\etad)\propto A_{\phid}\cos\phid + A_1\exp\left(-\frac{\phid^2}{2\sigma^2_{\phid}}-\frac{\etad^2}{2\sigma^2_{\etad}}\right).
\label{eq:minijet}
\ee
We note, however, that this decomposition is strongly model dependent. If the assumed functional form of Eq.~(\ref{eq:minijet}) is incorrect for minijet correlation, then the extracted minijet correlation will contain flow contributions, as discussed above. Therefore, the name ``minijet'' here should be taken merely as a label to refer to all the correlation structures except elliptic flow under the particular assumption of Eq.~(\ref{eq:minijet}). Keeping in mind the model dependence, we now proceed to estimate the magnitude of non-flow contributions from the decomposed minijet correlations.
%These correlations contribute to non-flow~\cite{trainor}. 

We shall assume minijet correlation does not vary with minijet direction relative to the reaction plane and particle emission within minijets is independent, and use Eq.~(\ref{eq16}) to estimate the magnitude of minijet contribution to non-flow from STAR's measurement~\cite{minijet}.
%The measured angular correlation (pair density per charged hadron) is parameterized by a two-dimensional Gaussian in $(\etad,\phid)$ and a dipole moment in $\dphi$:
%\be
%f(\phid,\etad)\propto A_{\phid}\cos\phid + A_1\exp\left(-\frac{\phid^2}{2\sigma^2_{\phid}}-\frac{\etad^2}{2\sigma^2_{\etad}}\right).
%\label{eq:minijet}
%\ee
From Eq.~(\ref{eq:minijet}), the average pair opening angle can be obtained by $\mean{\cphid}=\intpi f(\phid)\cphid d\phid$ and equals to $\mean{\cphi}^2_{cl}$ due to independent particle emission. Number of minijet particle-pairs per hadron ($V$) can be obtained by integrating $f(\phid,\etad)$ within the STAR acceptance (2 units of pseudo-rapidity $\eta$). Total number of minijet particle pairs within acceptance is then $\Ncl\Na^2=VN$ where $N=2\dNdeta$. To obtain the number of minijet particles per hadron $\Ncl\Na/N$, we note that the measured number of particle pairs per minijet is $VN/\Ncl$ where $\Ncl$ is number of clusters contributing to the measured signal, and in turn number of fragments per hadron is $\Ncl\Na/N=\Ncl\sqrt{VN/\Ncl}/N=\sqrt{V\Ncl/N}$. Number of minijets per unit of pseudo-rapidity should be just that in $pp$ scaled by the number of binary collisions, $0.013\Nbin$. However, minijets outside acceptance also contribute to the measured signal, thus $\Ncl>0.013\Nbin$; perhaps $\Ncl=0.013\Nbin\times\kappa$ where $\kappa\sim2$ given the measured broad $\eta$-width of minijet correlation. (Note, minijets inside acceptance always contribute to the measured signal, although some of the minijet fragments will leak out of acceptance, which makes the measured fragment multiplicity per minijet smaller.) Number of fragments per hadron is therefore $\Ncl\Na/N>\sqrt{0.013\Nbin V/(\dNdeta)}$, good to a factor of $\sqrt{\kappa}$. The cluster size (fragment multiplicity per minijet) measured inside acceptance is in turn $\Na=V/(\Ncl\Na/N)<\sqrt{V(\dNdeta)/(0.013\Nbin)}$, again good to a factor of $\sqrt{\kappa}$. In the estimates below, we assume $\kappa=1$.

In order to estimate non-flow effects arising from cross-talk between cluster correlation and cluster flow, terms $X$ and $Y$ in Eq.~(\ref{eq:X}) and (\ref{eq:Y}), respectively, 
%using Eq.~(\ref{eq16}), we also need to know the elliptic flow of clusters. 
we need to know the elliptic flow of clusters. 
This is not measured. We may estimate its magnitude at least in two ways, and they can give very different results. 
(i) Minijet cluster size (not for clusters measured in acceptance, but those in full space) is $\sqrt{V(\dNdeta)/(0.013\Nbin)}$ which is roughly 10 charged hadrons in central Au+Au collisions, suggesting the parent $\pt\sim6$~GeV/$c$. Those initial partons may have large $v_2$ (taking on the saturated value at high $\pt$), which is $\times$3--4 higher than the average. 
(ii) If clusters result from initial state (semi-)hard parton scatterings, then their distribution should be isotropic, $v_2=0$. Subsequent jet quenching results in cluster size varying with the reaction plane, larger out-of-plane than in-plane. This yields effectively a finite cluster $v_2$ (weighted by the number of particle pairs per cluster) which can be negative.
Thus, depending on physics scenario, cluster $v_2$ can be positive or negative. However, non-flow effects from cross-talk terms are generally small, significantly smaller than that from particle correlations within the same cluster, i.e., $C$ in Eq.~(\ref{eq:C}). Thus, we shall neglect non-flow effects from cross-talk in the following estimate. Again, it is worth to note that non-flow due to minijet-hydro and minijet-minijet correlations have identical angular shape as elliptic flow, so they cannot be distinguished in experiment and thus are included in essentially all elliptic flow measurements.

We estimate non-flow effect for top 5\% central Au+Au collisions at 200 GeV. The measured minijet amplitude is $A_1\approx0.63$, $\eta$-width is $\sigma_{\etad}\approx2.1$, and total extrapolated volume is $2\pi A_1\sigma_{\etad}\sigma_{\phid}\approx5.3$~\cite{minijet}. From these measurements we calculate the minijet $\phi$-width to be $\sigma_{\phid}\approx0.64$, the minijet azimuthal angle spread to be $\mean{\cphid}=\mean{\cphi}^2_{cl}\approx0.44$, and the minijet signal volume within STAR acceptance to be $V\approx1.9$. The number of minijet fragments is $\Ncl\Na/N\approx\sqrt{0.013\Nbin V/(\dNdeta)}\approx0.19$ where $\dNdeta=691$ and $\Nbin=1012$~\cite{Levente}. Non-flow contribution from minijet correlation is given by $\sum C=(V/N)\mean{\cphid}\approx0.025$. Comparison to the measured $\vv{2}=0.024$ by the two-particle method~\cite{v4} suggests that non-flow from correlation between particles from the same minijet cluster is the dominate contributor to $\vv{2}$.

We note, however, that the above estimate and conclusion are based on the assumption that Eq.~(\ref{eq:minijet}) is the proper functional form for minijet correlation so that the decomposition of the measured two-particle correlation into minijet and elliptic flow in Ref.~\cite{minijet} is correct. 
It is possible that our result may be an overestimate of non-flow contributions because, as discussed earlier, the minijet correlation may contain flow contributions and is most likely affected by hydrodynamic flow. In fact, ample evidence indicates that flow fluctuation effects are significant in central Au+Au collisions~\cite{Voloshin_fluc,PHOBOS}. The fact that our estimated non-flow contribution from minijet correlations dominates the measured $\vv{2}$ may suggest that the decomposition of elliptic flow and minijet correlations by Eq.~(\ref{eq:minijet}) is improper resulting in an overestimate of the non-flow.

\section{Summary}

We have derived analytical forms for non-flow contributions from cluster correlations to two-particle elliptic flow measure $\vv{2}$. We estimate non-flow contribution from two-body $\decay$ decays. With transverse distribution in accordance with measurement, non-flow contribution from $\rho$ decays is negative, contradictory to common perception. The magnitude of the non-flow contribution is small, on the order of a few percent for most centrality bins of Au+Au collisions. The large difference between two- and four-particle elliptic flow measurements cannot be due to resonance decays. The likely sources for the difference are eccentricity fluctuations and other non-flow effects such as jet correlations.

STAR has decomposed the two-particle $(\etad,\phid)$ correlation into a quadrupole component $\vv{2D}$ and the so-called minijet correlations assuming the functional form of Eq.~(\ref{eq:minijet}) for the latter~\cite{minijet}. We outlined a procedure to estimate the non-flow contributions from the decomposed minijet correlations to elliptic flow $\vv{2}$ measurement by the two-particle method. We estimated the magnitude of the non-flow contributions in central Au+Au collisions, and found the non-flow contributions from the decomposed minijet correlations to be predominant. This may suggest, because flow fluctuation effects are expected to be sizeable in central Au+Au collisions, that the assumed functional form for minijet correlations used in the decomposition may be improper.

Nevertheless, given the decomposed minijet cluster correlation, we found the main non-flow contribution is from correlations between particles within the same cluster. The magnitudes of non-flow contributions due to correlations between cluster particle and hydro-particle and between particles from different clusters are generally small, depending on the relative $v_2$ of hydro-particles and clusters themselves. Those non-flow effects from cross-talk of particles are included in all available elliptic flow measurements. Thus comparisons between measurements and hydro calculations should be taken with caution.

%Subtraction of flow background is one of the most important aspects of jet-correlation measurements. We argue that the elliptic flow value to be used in background subtraction should be $\vv{2D}$ fit to two-particle correlation data which excludes non-flow contribution from particle correlation within clusters. We defer this discussion on jet-correlation flow background to a future work.

\section*{Acknowledgment}

%We thank Dr.~Art Poskanzer, Dr.~Aihong Tang, Dr.~Tom Trainor, and Dr.~Sergei Voloshin for helpful discussions. 
This work is supported by U.S. Department of Energy under Grant DE-FG02-88ER40412.

%%%%%%%%%%%%%%%%%%%%%%%%%%%%%%%%%%%%%%%%%%%%%%%%%%%%%%%%%%%%%%%%%%%%%%


\begin{thebibliography}{99}
\bibitem{flow} K.~H.~Ackermann {\it et al.}~(STAR Collaboration), Phys.~Rev.~Lett.~{\bf 86}, 402 (2001).

\bibitem{whitepaper} J.~Adams {\it et al.}~(STAR Collaboration), Nucl.~Phys.~{\bf A757}, 102 (2005).

\bibitem{nonflow} J.~Adams {\it et al.}~(STAR Collaboration), Phys.~Rev.~Lett.~{\bf 93}, 252301 (2004).

\bibitem{v2method} A.~M.~Poskanzer and S.~A.~Voloshin, Phys.~Rev.~C {\bf 58}, 1671 (1998).

\bibitem{v4} C.~Adler {\it et al.}~(STAR Collaboration), Phys.~Rev.~C {\bf 66}, 034904 (2002).

\bibitem{Voloshin_fluc} S.~A.~Voloshin {\it et al.}, Phys.~Lett.~{\bf B659}, 537 (2008). %arXiv:0708.0800 [nucl-th]

\bibitem{PHOBOS} B.~Alver {\it et al.}~(PHOBOS Collaboration), Phys.~Rev.~C {\bf 81}, 034915 (2010). %arXiv:1002.0534 [nucl-ex]

\bibitem{note_Poisson} This is because the product of two Poisson variables is not Poisson. Note that, in real data analysis of heavy-ion collisions where multiplicity windows are often applied, none of the multiplicities may be Poisson.

\bibitem{note} We have examined non-uniform decay angle distributions due to possible polarization effect, and found they do not change our results qualitatively. 

\bibitem{rho} J.~Adams {\it et al.}~(STAR Collaboration), Phys.~Rev.~Lett.~{\bf 92}, 092301 (2004).

\bibitem{v2data} J.~Adams {\it et al.}~(STAR Collaboration), Phys.~Rev.~C {\bf 72}, 014904 (2005).

\bibitem{spectra} J.~Adams {\it et al.}~(STAR Collaboration), Phys.~Rev.~Lett.~{\bf 92}, 112301 (2004).

\bibitem{kstar} C.~Adler {\it et al.}~(STAR Collaboration), Phys.~Rev.~C {\bf 66}, 061901 (2002).

\bibitem{Levente} B.~I.~Abelev {\it et al.}~(STAR Collaboration), Phys.~Rev.~C {\bf 79}, 034909 (2009). %arXiv:0808.2041.

\bibitem{minijet} M.~Daugherity (STAR Collaboration), J.~Phys.~G {\bf 35}, 104090 (2008). %arXiv:0806.2121.

\bibitem{trainor} T.~A.~Trainor, Phys.~Rev.~C {\bf 78}, 064908 (2008). %arXiv:0803.4002.

\end{thebibliography}
\end{document}